\documentclass[12pt]{iopart}

\usepackage{iopams}
\usepackage{graphicx}
\usepackage{subfigure}

\newcommand{\text}[1]{\ensuremath{{\rm #1}}}

\newcommand{\rp}{\ensuremath{P_\text{R}}}

\newcommand{\rpf}{\ensuremath{\tilde{P}_\text{R}}}

\newcommand{\scgap}{\ensuremath{\Delta_\text{g}}}

\newcommand{\zscgap}{\ensuremath{\Delta_\text{0}}}

\newcommand{\hf}[1]{\ensuremath{q_{\,\text{#1}}}}

\newcommand{\gmat}{\ensuremath{\mathsf{G}_0}}

\newcommand{\cmat}{\ensuremath{\mathsf{C}_0}}

\newcommand{\rmat}{\ensuremath{\mathsf{R}_0}}

\newcommand{\smat}{\ensuremath{\mathsf{S}_0}}

\newcommand{\kmat}{\ensuremath{\mathsf{K}_0}}

\begin{document}

\title{Electrothermal Model of Kinetic Inductance Detectors}
\author{C.N. Thomas, S. Withington and D.J. Goldie}
\address{Cavendish Laboratory, JJ Thomson Avenue, Cambridge, CB3 0HE, UK}
\ead{c.thomas@mrao.cam.ac.uk}

\begin{abstract}
An electrothermal model of Kinetic Inductance Detectors (KIDs) is described. The non-equilibrium state of the resonator's quasiparticle system is characterized by an effective temperature, which because of readout-power heating is higher than that of the bath.
By balancing the flow of energy into the quasiparticle system, it is possible to calculate the steady-state large-signal, small-signal and noise behaviour.
Resonance-curve distortion and hysteretic switching appear naturally within the framework.
It is shown that an electrothermal feedback process exists, which affects all aspects of behaviour.
It is also shown that generation-recombination noise can be interpreted in terms of the thermal fluctuation noise in the effective thermal conductance that links the quasiparticle and phonon systems of the resonator.
Because the scheme is based on electrothermal considerations, multiple elements can be added to simulate the behaviour of complex devices, such as resonators on membranes, again taking into account readout power heating.
\end{abstract}

\maketitle


\section{Introduction}\label{sec:introduction}

Kinetic Inductance Detectors (KIDs)\,\cite{mazin2006position, o2011arcons, day2006antenna,moore2012position} comprise a superconducting thin-film microwave resonator connected to an optical or submillimetre-wave absorber of some kind.
The resonator operates at a temperature $T$ well below the superconducting transition temperature $T_\text{C}$: $T / T_\text{C} \approx 0.1$.
In this regime, the resonant frequency is strongly influenced by the surface kinetic inductance of the film, which changes when an energetic photon or phonon is absorbed.
The design is intrinsically frequency-multiplexable, as multiple devices tuned to different resonant frequencies can be coupled to the same readout transmission line\,\cite{day2003broadband}. 
The devices can then be readout simultaneously by using a software-defined radio receiver to probe the changes in transmission gain and phase shift along the line in response to illumination.
KIDs are being developed for applications across the whole of the electromagnetic spectrum, with particular emphasis on realizing large-format imaging arrays and chip spectrometers\,\cite{endo2013chip, shirokoff2012mkid, thomas2013cambridge}.

There is increasing evidence that the energy dissipated in the resonator by the microwave readout signal places fundamental limits on device performance.
Although KIDs are read out at frequencies less than 10\,GHz, well-below the direct pair breaking frequency of most low-$T_\text{C}$ superconductors (e.g. 160\,GHz for Al at $T / T_\text{C} = 0.1$), the readout signal can still break pairs through multiple-photon absorption events.
The non-thermal quasiparticle distributions that result have been studied by Goldie\,\cite{goldie2013non}, and experimental evidence of these non-equilibrium states has been reported by de Visser\,\cite{de2014evidence}.
At low bath temperatures, the excess quasiparticles generated by the readout signal dominate the thermal background, determine the dynamical behaviour, and limit the sensitivity due to increased generation-recombination noise\,\cite{de2012generation}.
de Visser\,\cite{de2010readout} and Thompson\,\cite{thompson2013dynamical} have further shown that the effective `heating' of the quasiparticle system can lead to hysteretic switching, which limits the maximum readout power that can be applied, and so impacts sensitivity and dynamic range.

In addition to this unavoidable heating, designs are now emerging where parts, or all, of the resonator are suspended on a dielectric membrane\,\cite{quaranta2013x,timofeev2014submillimeter,lindeman2014arrays}.
The aim is to increase phonon-trapping in the superconducting film, and thereby increase quasiparticle generation efficiencies and relaxation times, or to improve the spatial uniformity of the optical response of the detector.
In all cases, the thermal operating point is affected by dissipated readout power elevating the temperature of the membrane above that of the bath.
There is also a dynamical electrothermal feedback effect, because as a system moves out of equilibrium, e.g. due to a change in optical signal power, the resonant frequency changes, the coupling of readout power into the device changes, and the readout-power heating changes.
This feedback can improve the device's operating characteristics or make them worse depending on the design.

We present a steady-state electrothermal model that takes into account quasiparticle and substrate heating in KIDs.
We formulate the theory in a completely general way so that it can be applied to complex devices having multiple thermal elements.
The absorption of readout power establishes a quiescent effective quasiparticle temperature, which is above that of the bath.
The behaviour is then linearised about this operating point to determine the small-signal and noise characteristics.
The model takes into account the dynamical feedback processes associated with readout-power heating.
Our electrothermal model reduces to the commonly used quasiparticle-number model in the appropriate limits.

\section{Thermal model}\label{sec:thermal_model}

Goldie\,\cite{goldie2013non} has investigated the non-equilibrium distributions of quasiparticles and phonons in a superconductor in the presence of sub-gap microwave readout power.
The simulations were performed within the rate-equation framework of Chang and Scalapino\,\cite{chang1977kinetic}.
Goldie's simulations result in a number of important observations that motivate an electrothermal model that can be used for device-level calculations: (i) The driven quasiparticle distribution is well-approximated globally by a thermal distribution having an effective temperature $T_\text{qp}$, which is higher than the effective phonon temperature $T_\text{ph}$.
The non-equilibrium values of quantities such as surface impedance can be calculated to good accuracy using equilibrium expressions and the effective quasiparticle temperature.
(ii) The energy flow between the quasiparticle and phonon systems is well described by a simple expression involving $T_\text{qp}$ and $T_\text{ph}$, in an analogous manner to low-temperature electron-phonon coupling in normal metals\,\cite{goldie2013non}.
(iii) The original modelling has been extended to the case where above-gap signal photons and sub-gap readout photons are absorbed simultaneously, and to a range of superconducting materials \,\cite{guruswamy2014non}.
This work shows that although there are subtle state-blocking effects, it is sufficient for most purposes to describe the effects of the readout and signal photon fluxes as heating processes.

\begin{figure}
\centering
{\includegraphics[width=7cm]{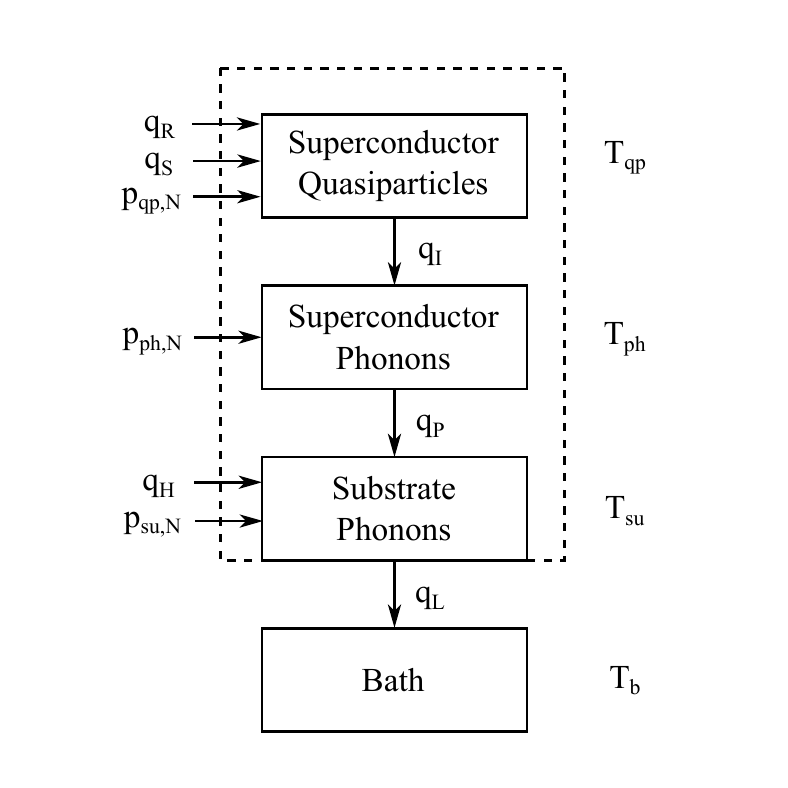}}
\label{fig:thermal_model}
\caption{Electrothermal model of a generic KID.
$T_\text{qp}$, $T_\text{ph}$, $T_\text{su}$ and $T_\text{b}$ are the effective quasiparticle, phonon, substrate and bath temperatures respectively.
$\hf{R}$ and $\hf{S}$ denote the dissipated read-out and signal powers respectively.
$\hf{H}$ represents any power dissipated in the substrate directly, say by a calibration heater.
$\hf{I}$, $\hf{P}$ and $\hf{L}$ are internal power flows.
The terms of the form $p_\text{\dots,N}$  are effective noise sources.}
\end{figure}

\begin{figure}
\centering
\includegraphics[width=7cm]{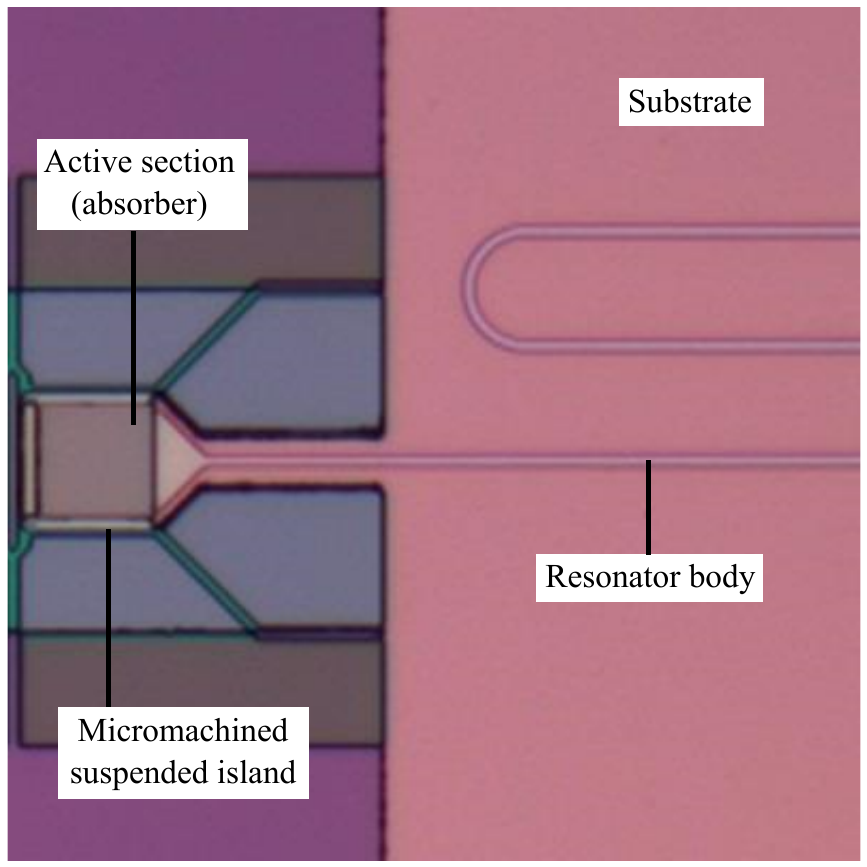}
\label{fig:kid_photo}
\caption{SiN membrane-supported optical-photon counting KID fabricated in the Physics Department, University of Cambridge.}
\end{figure}

On the basis of these observations, it is possible to create an electrothermal model that takes into account quasiparticle heating, and that can be used to explore the behaviour of membrane-supported devices.
Figure \ref{fig:thermal_model} shows a generic model of a KID.
It distinguishes the quasiparticle system of the resonator, the phonons in the resonator, the phonon system of the substrate, which may be a suspended membrane, and the phonon system of the bath.
$\hf{S}$ is the rate at which energy is supplied by a pair-breaking optical signal, $\hf{R}$ is the readout power dissipated in the resonator, and $\hf{H}$ represents any thermal power applied to the substrate directly: say from a calibration resistor or microstrip RF load on a membrane.
The arrows indicate assumed directions of energy flow.
Models of this kind can be used to represent the behaviour of all devices encountered in practice: for example, the membrane-supported optical KID shown in Figure \ref{fig:kid_photo}.
In fact when testing devices of the kind shown in Figure \ref{fig:kid_photo}, we have found that hysteretic switching occurs at readout powers of order 100 times smaller than identical substrate-mounted devices, indicating that strong electrothermal feedback processes are present. 
In the case of KIDs on bulk substrates, it is only necessary to use the elements in the dashed box.

Assuming the resonator can be modelled as an impedance shunting a lossless readout line, the dissipated power is given by
\begin{equation}\label{eqn:readout_power_dissipation}
	\hf{R} = 2 \Re \left[ S_{21}^{}(\nu_\text{r})
		\left( 1 - S_{21}^*(\nu_\text{r}) \right) \right] \rp,
\end{equation}
where $S_{21}$ is the forward $S$-parameter of the resonator defined using reference planes on the readout line, $\nu_\text{r}$ is the readout frequency, $\rp$ is the incident readout power, and $\Re$ denotes the real part.
The calculation of $S_{21}$ is discussed in Section \ref{sec:resonator_steady_state}.
An important question is into which block in Figure \ref{fig:thermal_model} does $\hf{R}$ flow?
Ohmic dissipation transfers energy to the quasiparticle system, whereas dielectric or Two-Level-System (TLS) loss\,\cite{gao2008semiempirical} transfers energy to substrate phonons.
In the case of a transmission-line, these mechanisms can be described by a distributed series resistance and a shunt conductance respectively, giving attenuation coefficients $\alpha_1$ and $\alpha_2$.
It is then straightforward to show that the total power dissipated divides according to the ratios $\alpha_1 / (\alpha_1 + \alpha_2)$ and $\alpha_2 / (\alpha_1 + \alpha_2)$, allowing $\hf{R}$ to be partitioned
between the quasiparticle and phonon blocks in Figure \ref{fig:thermal_model}.

The power dissipated by a pair-breaking optical signal is
\begin{equation}\label{eqn:optical_heat_flow}
	\hf{S} = \eta P_\text{S},
\end{equation}
where $P_\text{S}$ is the incident optical power and $\eta$ is the optical efficiency.
Other models of thin-film superconducting devices normally include a quasiparticle generation efficiency in the optical term\,\cite{zmuidzinas2012superconducting,guruswamy2014non}.
This accounts for the fraction of energy lost on short time scales when high-energy (potentially pair-breaking) phonons escape from the film. 
However, this behaviour is included explicitly in our model via the heat flow between the superconductor and substrate phonon systems.

The rate of energy flow between the quasiparticle and phonon systems, $\hf{I}$, can be approximated by
\begin{eqnarray}
\label{eqn:qp_ph_heat_flow}
	\hf{I}(T_\text{qp}, T_\text{ph})
	=& \, \frac{V \Sigma_\text{s}}{\eta_{2 \scgap(P_\text{abs})}} \left[ T_\text{qp}
		\exp \left( -2 \scgap (T_\text{qp}) / k_\text{b} T_\text{qp} \right) \right. \nonumber \\
	&- \left. T_\text{ph} \exp \left( -2 \scgap (T_\text{ph})/ k_\text{b} T_\text{ph} \right)
	\right],
\end{eqnarray}
where $V$ is the active volume of the resonator, $T_\text{ph}$ is the phonon temperature, $\Sigma_\text{s}$ is a material-dependent constant and $\eta_{2 \scgap(P_\text{abs})}$ is the fraction of $q_\text{I}$ resulting from recombination of quasiparticles into Cooper pairs\,{\cite{goldie2013non}}.
In what follows, all temperatures are effective temperatures unless otherwise stated.
Strictly, (\ref{eqn:qp_ph_heat_flow}) is valid for $T_\text{qp} / T_\text{C} < 0.5$ and we have we have derived the appropriate coefficients for a variety of materials in this regime\,\cite{guruswamy2014non}.
The physical origin of (\ref{eqn:qp_ph_heat_flow}) is discussed in \ref{sec:qp_ph_heat_flow}. 

(\ref{eqn:qp_ph_heat_flow}) appears in a slightly different form in Goldie\,\cite{goldie2013non} and Thompson\,\cite{thompson2013dynamical}.
Firstly, the dependence was on the bath temperature $T_\text{b}$ rather than the superconductor phonon temperature $T_\text{ph}$.
This was a convenience introduced in the original papers because it was found that $T_\text{ph} \approx T_\text{b}$.
Secondly, in \cite{goldie2013non} and \cite{thompson2013dynamical}, (\ref{eqn:qp_ph_heat_flow}) includes an additional factor of $[1 + \tau_\text{l} / \tau_\text{pb}]^{-1}$ that accounts for the loss of pair-breaking phonons from the film into the substrate; $\tau_\text{l}$ and $\tau_\text{pb}$ are, respectively, the characteristic timescales over which loss and pair-breaking processes occur.
In this paper we represent phonon-loss by an explicit heat flow term between the phonon systems of the superconductor and substrate, so the factor is unnecessary.
This partitioning allows the inclusion of the effects of the superconductor phonon heat capacity in the small signal model, which is not possible in the original formulation.

The heat flow between the phonon system of the superconductor and the phonon system of the substrate, $\hf{P}$, is given by
\begin{equation}
\label{eqn:ph_ph_heat_flow}
	\hf{P} (T_\text{ph}, T_\text{su})
	= A \Sigma \bigl[ T_\text{ph}^n - T_\text{su}^n \bigr],
\end{equation}
where $T_\text{su}$ is the temperature of the substrate phonons, $A$ is the area of the interface, and $\Sigma$ is a material-dependent parameter.
It is common to take $n=4$ on the basis of the acoustic- and diffuse-mismatch models of phonon scattering at a boundary\,\cite{swartz1989thermal}.

When a suspended membrane is used, the heat flow from the substrate to the bath, $q_{\text{L}}$, has a similar form to (\ref{eqn:ph_ph_heat_flow}), but $n$ can take on a range of values depending on the cross-section and length of the legs used to support the membrane.
Calculations of this kind are discussed in \cite{rostem2008technique} and \cite{withington2011low}, and are well known in bolometer theory.

While we have assigned the superconductor phonons a single effective temperature, models also exist in which the phonons are divided into two populations, with different effective temperatures, based on their ability to break Cooper pairs\,\cite{parker1975modified}.
This and other more detailed models can be treated within the framework of this paper by modification of the underlying thermal model. 
For example, the two-temperature models just described can be incorporated by splitting the superconductor phonons in Figure \ref{fig:thermal_model} into two heat capacities representing the sub-populations, then partitioning the heat flows accordingly.

\section{Microwave response}\label{sec:resonator_and_readout}

In our scheme, the state of a resonator is parameterized by the effective quasiparticle temperature, $T_\text{qp}$, and therefore it is necessary to consider how perturbative changes in $T_\text{qp}$ about some quiescent value lead to small changes in the amplitude and phase of the transmitted microwave signal.
In what follows, we shall use $\tilde{g}(\nu)$ to denote the one-sided spectral representation of a time-domain signal $g(t)$:
\begin{equation}\label{eqn:spectral_representation}
	 g(t) = \int_0^\infty \Re \left[ \tilde{g}(\nu) e^{2 \pi i \nu t} \right] \, d\nu.
\end{equation}
$\nu$ will be used for frequencies near the readout frequency $\nu_\text{r}$, and $f$ will be used for signal frequencies parametrically down-converted into the range $0 \leq f \ll \nu_\text{r}$.

\subsection{Resonator behaviour}\label{sec:resonator_steady_state}

\begin{figure}
\centering
\includegraphics[width=8cm]{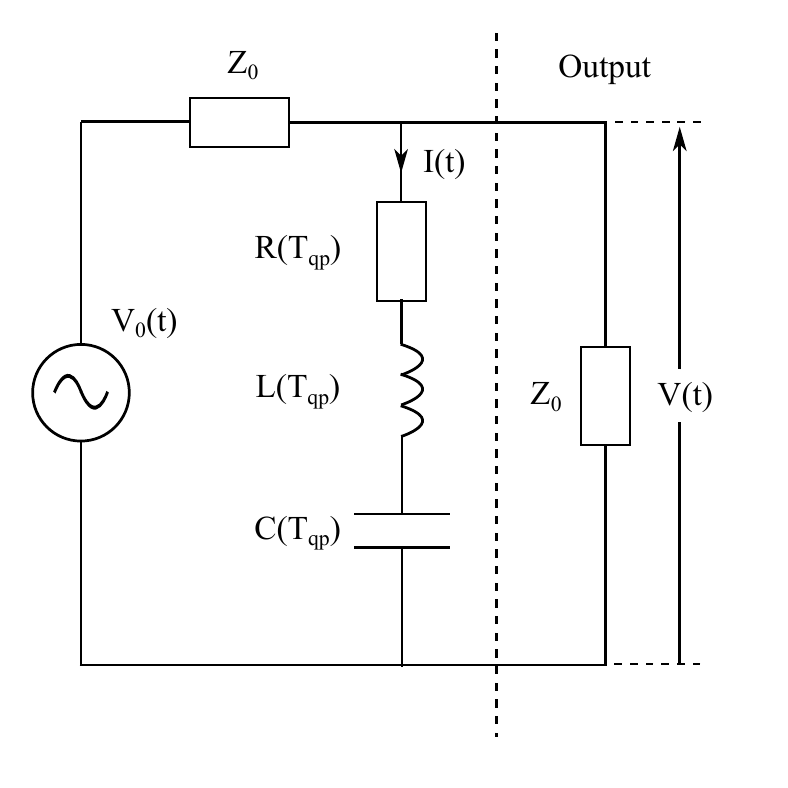}
\caption{\label{fig:equivalent_circuit} Equivalent circuit of a KID resonator.}
\end{figure}

To calculate microwave response, a KID can be modelled in an architecture-independent manner as a series RLC resonator shunting a readout line: Figure \ref{fig:equivalent_circuit}.
$V_0(t)$ represents the applied readout voltage, $Z_0$ the characteristic impedance of the readout line and $V(t)$ the transmitted voltage.
$R$, $L$ and $C$ are the effective lumped resistance, inductance and capacitance of the KID which, for generality, are all assumed to depend on $T_\text{qp}$.
Analysing this circuit, we find $2V(t) = V_0(t) - Z_0 I(t) $ and
\begin{eqnarray}\label{eqn:resonator_de}
	 \hat{F} (T_\text{qp}) I(t) = \partial_t V_0(t),
\end{eqnarray}
where the differential operator $\hat{F}$ is defined as
\begin{eqnarray}\label{eqn:def_g_operator}
	\hat{F}(T_\text{qp}) =
		2 L(T_\text{qp}) \, \partial_t^2
		+ \bigl[ 2 R(T_\text{qp}) + Z_0 \bigr] \, \partial_t
		+ 2 C(T_\text{qp})^{-1}.
\end{eqnarray}
Strictly (\ref{eqn:resonator_de}) and (\ref{eqn:def_g_operator}) assume that the rate of change of $T_\text{qp}$ is smaller than the rates of change of $V$ and $I$, so that $L$ and $R$ can be extracted from the time derivatives.
However, this requirement is always met under normal operating conditions.
For constant $T_\text{qp}$, the solution of (\ref{eqn:resonator_de}) in the spectral domain is
\begin{eqnarray}
	\tilde{V}(\nu) = \frac{1}{2} S_{21}(\nu, T_\text{qp}) \tilde{V}_0 (\nu)
	\label{eqn:ss_output_voltage} \\
	\tilde{I}(\nu) = -Z_0^{-1} S_{11}(\nu, T_\text{qp}) \tilde{V}_0 (\nu)
	\label{eqn:ss_resonator_current} \\
	S_{11}(\nu, T_\text{qp}) = -\frac{Q_\text{T}}{Q_\text{C}}
		\Biggl[ 1 + 2 i Q_\text{T} \frac{\nu - \nu_0}{\nu_0} \Biggr]^{-1}
	\label{eqn:resonator_s11} \\
	S_{21}(\nu, T_\text{qp}) = 1 + S_{11}(\nu, T_\text{qp})
	\label{eqn:resonator_s21},
\end{eqnarray}
where $\nu_0 = 1 / 2 \pi \sqrt{LC}$ is the resonant frequency, $Q_\text{C} = 4 \pi \nu_0 L / Z_0$ is the coupling quality factor, $Q_\text{I} = 2 \pi \nu_0 L / R$ is the internal quality factor and $Q_\text{T} = Q_\text{I} Q_\text{C} / (Q_\text{I} + Q_\text{C})$ is the total quality factor.	
$S_{11}$ and $S_{21}$ are the de-embedded values of the reflection and transmission $S$-parameters, as would be measured at the device plane.

Operationally, a homodyne readout scheme is usually used, in which a microwave tone is split between two signal paths, one through the KID and one bypassing it.
The bypass and through signals are then used to drive the LO and RF ports respectively of an I/Q mixer.
The in-phase ($\mathcal{I}$) and quadrature ($\mathcal{Q}$) outputs of the mixer can be expressed as
\begin{eqnarray}
	\mathcal{I}(t) = \int_{-\infty}^t
		R_\text{I} (t - t') V_\text{in} (t') \cos 2 \pi \nu_\text{r} t' \, dt'
		\label{eqn:t_domain_i} \\
	\mathcal{Q}(t) = -\int_{-\infty}^t
		R_\text{Q} (t - t') V_\text{in} (t') \sin 2 \pi \nu_\text{r} t' \, dt'
		\label{eqn:t_domain_q},
\end{eqnarray}
where $\nu_\text{r}$ is the readout frequency, $V_\text{in}(t)$ is the input voltage to the mixer, and $R_\text{I}(t)$ and $R_\text{Q}(t)$ are response functions that characterize post-conversion filtering.
$V_\text{in}(t)$ is related to $V(t)$, the voltage across the KID, by
\begin{equation}\label{eqn:t_domain_iq_in}
	V_\text{in}(t) = \int_{-\infty}^t R_0(t - t') V (t')\, dt',
\end{equation}
where the effects of amplification and cabling have been collected into a system response function $R_0(t)$.
In the remainder of the paper we assume that the post-conversion bandwidth is always small enough that $0 \leq f \ll \nu_\text{r}$, and then in the spectral domain (\ref{eqn:t_domain_i}) and (\ref{eqn:t_domain_q}) become
\begin{eqnarray}
\label{eqn:spec_domain_iq}
	\tilde{\mathcal{I}}(f) = + \frac{1}{8} \tilde{R}_\text{I} (f) & 
	\left[ \tilde{R}_0(\nu_\text{r} + f)\tilde{V}(\nu_\text{r} + f) \right. \nonumber \\
	& \left. + \tilde{R}_0^* (\nu_\text{r} - f) \tilde{V}^* (\nu_\text{r} - f) \right] \nonumber \\ 
	\tilde{\mathcal{Q}}(f) = - \frac{i}{8} \tilde{R}_\text{Q} (f) &
	\left[ \tilde{R}_0(\nu_\text{r} + f) \tilde{V}(\nu_\text{r} + f) \right. \nonumber \\
	& \left. - \tilde{R}_0^* (\nu_\text{r} - f) \tilde{V}^* (\nu_\text{r} - f) \right].
\end{eqnarray}
When the quasiparticle temperature is constant,  $T_\text{qp} = T_0$, (\ref{eqn:ss_output_voltage}) can be used in (\ref{eqn:spec_domain_iq}) to show
\begin{eqnarray}
	\tilde{\mathcal{I}}(0) = \frac{1}{8} \Re \Bigr[
		\tilde{R}_\text{I}(0) \tilde{R}_0(\nu_\text{r}) S_{21}(\nu_\text{r}, T_0) V_0 \Bigl]
		\label{eqn:static_I} \\
	\tilde{\mathcal{Q}}(0) = \frac{1}{8} \Im \Bigr[
		\tilde{R}_\text{Q}(0) \tilde{R}_0(\nu_\text{r}) S_{21}(\nu_\text{r}, T_0) V_0 \Bigl]
		\label{eqn:static_Q}.
\end{eqnarray}
The outputs of the I/Q mixer therefore measure the real and imaginary parts of the product of the transmission characteristic of the measurement system and the desired transmission characteristic, $S_{21}$, of the KID.

\subsection{Temperature perturbations}\label{sec:parametric_modulation}

Now assume that $T_\text{qp}$ is varied according to $T_\text{qp}(t) = T_0 + \Delta T_\text{qp}(t)$, where $T_0$ is the quiescent, or operating-point temperature of the quasiparticle system.
Techniques for finding $T_0$ will be discussed later.
The resonator current can be written $I_0(t) + \Delta I(t)$, where $I_0(t)$ is the current when $T_\text{qp} = T_0$, and $\Delta I(t)$ is the perturbation produced by $\Delta T_\text{qp} (t)$.
Because the temperature dependence is in the coefficients of (\ref{eqn:resonator_de}), the operator $\hat{F}(T_\text{qp}(t))$ may be written
\begin{equation}\label{eqn:perturbed_operator}
	\hat{F}(T_\text{qp}(t)) \approx \hat{F}(T_0)
	+ \Delta T_\text{qp}^{}(t)
	\cdot \bigl[ \partial_{\,T_\text{qp}} \, \hat{F} (T_\text{qp}) \bigr]_{T_0}.
\end{equation}
(\ref{eqn:perturbed_operator}) and the perturbed resonator current can then be substituted into (\ref{eqn:resonator_de}).
Cancelling steady state terms and neglecting those that are second order in the perturbations, we find that $\Delta I(t)$ obeys
\begin{equation}\label{eqn:delta_i_de}
	\hat{F}(T_0) \Delta I(t) =  -\Delta T_\text{qp} (t) \cdot
	\bigl[ \partial_{\,T_\text{qp}} \, \hat{F} (T_\text{qp}) \bigr]_{T_0} I_0(t),
\end{equation}
where we again emphasise that $\Delta T_\text{qp} (t)$ varies slowly with respect to the microwave voltages and currents.
(\ref{eqn:delta_i_de}) can be recognised as a variant of (\ref{eqn:resonator_de}), where the effective driving voltage is derived from $I_0(t)$.
Remembering that $I_0(t) = \tilde{I}_{0} \exp \left[ 2 \pi i \nu_\text{r} t \right]$, the right-hand side of (\ref{eqn:delta_i_de}) can be expressed as
\begin{eqnarray}\label{eqn:u_driving_term}
	-\frac{1}{2} \Re &\Biggl\{ 2 \pi i \nu_\text{r} V_0
		\bigl[1 - S_{21}(\nu_\text{r}, T_0) \bigr]^{-1}
		\bigl[ \partial_{\, T_\text{qp}} S_{21}^{} \bigr]_{\nu_\text{r}, T_0} \nonumber \\
		& \times \left[
		\int_{\nu_\text{r}}^\infty \Delta \tilde{T}_\text{qp}(\nu - \nu_\text{r})
		\, e^{2 \pi i \nu t} \, d\nu \nonumber \right. \\
		& +
		\left. \int_0^{\nu_\text{r}} \Delta \tilde{T}_\text{qp}^*(\nu_\text{r} - \nu)
		\, e^{2 \pi i \nu t} \, d\nu \right] \Biggr\},
\end{eqnarray}
where we have replaced $\Delta T_\text{qp}(t)$ with its spectral representation and made use of the fact $\partial_T [\hat{F}(T)]_{T_0} I(t, T_0) = - \hat{F}(T_0) \partial_T [I(t, T)]_{T_0}$ when $I(t, T)$ is a solution of (\ref{eqn:resonator_de}). 
In addition we have dropped a term that involves evaluating $\Delta \tilde{T}_\text{qp} (\nu)$ at frequencies $\nu > \nu_\text{r}$, in keeping with our assumption that $T_\text{qp}$ varies more slowly than the current.
It is then straightforward to solve (\ref{eqn:delta_i_de}) in the spectral domain using (\ref{eqn:u_driving_term}), giving
\begin{eqnarray}\label{eqn:u_pertubration_current}
	&\Delta \tilde{I} (\nu) = - \frac{V_0}{2 Z_0} \frac{\nu_\text{r}}{\nu}
	\bigl[1 \! - \! S_{21}(\nu_\text{r}, T_0) \bigr]^{-1}
	\bigl[1 \! - \! S_{21}(\nu, T_0) \bigr] \nonumber \\
	&\times \bigl[ \partial_{\, T_\text{qp}} S_{21}^{} \bigr]_{\nu_\text{r},T_0}
	\cases{
 		\Delta \tilde{T}^*_\text{qp} (\nu_\text{r} - \nu) & for $\nu < \nu_\text{r}$ \\
		\Delta \tilde{T}^{}_\text{qp} (\nu - \nu_\text{r}) & for $\nu > \nu_\text{r}$
	}.
\end{eqnarray}
(\ref{eqn:u_pertubration_current}) shows that the spectrum of changes in the temperature appears in the resonator current in two sidebands around $\nu_\text{r}$: i.e. the device functions as a parametric upconverter.

It follows from $2V(t) = V_0 (t) - Z_0 I(t)$ that the current perturbation $\Delta I(t)$ produces an additive change of $\Delta V(t) = -Z_0 \Delta I (t) / 2$ in the output voltage $V(t)$.
(\ref{eqn:spec_domain_iq}) can then be used to be used to determine the corresponding changes $\Delta \mathcal{I}$ and $\Delta \mathcal{Q}$ in the outputs of the I/Q mixer:
\begin{equation}
\label{eqn:total_parameter_response}
	\left\{ \begin{array}{c}
		\Delta \tilde{\mathcal{I}}(f) \\
		\Delta \tilde{\mathcal{Q}}(f)
	\end{array} \right\}
	=
	\left\{ \begin{array}{c}
		\tilde{F}_\text{I} (f) \\
		\tilde{F}_\text{Q} (f)
	\end{array} \right\}
	\cdot
	\Delta \tilde{T}_\text{qp} (f),
\end{equation}
where
\begin{eqnarray}\label{eqn:parametric_response_function}
	\left\{ \begin{array}{c}
		\tilde{F}_\text{I} (f) \\
		\tilde{F}_\text{Q} (f)
	\end{array} \right\}
	= \, \frac{1}{32}
	\left\{ \begin{array}{c}
		\! \tilde{R}_\text{I} (f) \! \\
		\! -i \tilde{R}_\text{Q} (f)
	\end{array} \right\} \nonumber \\
	\times \left[ \,
	\frac{V_0^{} \tilde{R}_0^{} (\nu_\text{r} + f)
	\bigl[ 1 - S_{21}(\nu_\text{r} + f, T_0) \bigr] }
	{1 - S_{21}^{}(\nu_\text{r}, T_0)}
	\bigl[ \partial_{\, T_\text{qp}} S_{21}^{} \bigr]_{\nu_\text{r}, T_0}
	\right. \nonumber \\
	\left. + \left\{ \begin{array}{c}
		\! + \! \\
		\! - \!
	\end{array} \right\}
	\frac{V_0^{*} \tilde{R}_0^{*} (\nu_\text{r} - f)
	\bigl[ 1 - S_{21}^*(\nu_\text{r} - f, T_0) \bigr] }
	{1 - S_{21}^*(\nu_\text{r}, T_0)}
	\bigl[ \partial_{\, T_\text{qp}} S_{21}^{*} \bigr]_{\nu_\text{r},T_0}
	\right].
\end{eqnarray}
It has been assumed that $f \ll \nu_\text{r}$, so that $\nu_\text{r} / \nu \approx 1$.
According to (\ref{eqn:total_parameter_response}), the outputs of the mixer contain spectral information about the variation in quasiparticle temperature.
The transfer function $\tilde{F}(f)$ takes into account (i) parametric upconversion of the variations into sidebands of the readout frequency through the modulation of the resonance feature, as described by the terms involving $S_{21}$, (ii) the passage to the mixer at microwave frequencies through $\tilde{R}_0 (\nu)$, and (iii) the subsequent down-conversion in the mixer through $\tilde{R}_\text{I} (\nu)$ and $\tilde{R}_\text{Q} (\nu)$.
Thus, if the optical power, $\hf{S}$, changes, the thermal circuit can be analysed to give the change in $T_\text{qp}$, which in turn gives the I/Q outputs through (\ref{eqn:total_parameter_response}).

\subsection{Readout-power modulation}\label{sec:readout_tone_modulation}

A change in applied readout power is unique in that it affects the outputs of the I/Q mixer in two ways: (i) there is a differential change in quasiparticle temperature through heating; (ii) the transmitted output changes directly.
(\ref{eqn:total_parameter_response}) accounts for the former, but it is necessary to derive a separate expression for the latter.

Assume that the readout tone is amplitude modulated by a function $m(t)$:
\begin{equation}
	V_0(t) = \bigl[ 1 + m(t) \bigr] \Re \bigl[ V_0 e^{2 \pi i \nu_\text{r} t} \bigr],
\end{equation}
where the bandwidth of $m(t)$ is smaller than $\nu_\text{r}$.
This modulation can be represented by adding a term $\Delta V_0(t)$ to the readout signal having the spectrum
\begin{equation}
\label{eqn:modulation_term}
	\Delta \tilde{V}_0 (\nu) = \frac{V_0}{2}
	\cases{
		\tilde{m}^*(\nu_\text{r} - \nu) & for $\nu < \nu_\text{r}$ (LSB) \\
		\tilde{m}^{}(\nu - \nu_\text{r}) & for $\nu \geq \nu_\text{r}$ (USB),
	}.
\end{equation}
(\ref{eqn:ss_output_voltage}) and (\ref{eqn:spec_domain_iq}) can now be used to calculate the corresponding change in the resonator output voltage and I/Q mixer outputs respectively:
\begin{equation}\label{eqn:total_modulation_response}
	\left\{ \begin{array}{c}
		\Delta \tilde{\mathcal{I}}(f) \\
		\Delta \tilde{\mathcal{Q}}(f)
	\end{array} \right\}
	=
	\left\{ \begin{array}{c}
		\tilde{H}_\text{I} (f) \\
		\tilde{H}_\text{Q} (f)
	\end{array} \right\}
	\, \tilde{m}(f),	
\end{equation}
where the transfer functions are given by
\begin{eqnarray}
	& \left\{ \begin{array}{c}
		\tilde{H}_\text{I} (f) \\
		\tilde{H}_\text{Q} (f)
	\end{array} \right\}
	= \, \frac{1}{32}
	\left\{ \begin{array}{c}
		\! \tilde{R}_\text{I} (f) \! \\
		\! -i \tilde{R}_\text{Q} (f) \!
	\end{array} \right\} \nonumber \\
	& \times \Biggl[ \,
	V_0^{} \tilde{R}_0^{} (\nu_\text{r} + f) S_{21}^{}(\nu_\text{r} + f, T_0) \nonumber \\
	&+ \Biggl\{ \begin{array}{c}
		\! + \!  \\
		\! - \!
	\end{array} \Biggr\}
	V_0^{*} \tilde{R}_0^{*} (\nu_\text{r} - f) S_{21}^{*}(\nu_\text{r} - f, T_0)
	\, \Biggr].
\end{eqnarray}
For comparison with experiment, it is more convenient to work in terms of the change in incident readout power $\Delta \rp$ rather than the modulation function.
The two are related by
\begin{equation}\label{eqn:modulation}
	\Delta \rpf (f) = 2 \rp \tilde{m}(f),
\end{equation}
where $\rp$ is the quiescent readout power applied to the KID.

\subsection{Combined response}\label{sec:overall_response}

Combining the results of Sections \ref{sec:parametric_modulation} and \ref{sec:readout_tone_modulation}, the overall response at the outputs of the I/Q mixer are
\begin{eqnarray}
\label{eqn:temperature_responsivity}
\Delta \tilde{\mathcal{I}}(f) & = &
		\tilde{F}_\text{I} (f) \, \Delta \tilde{T}_\text{qp} (f)
		+ \tilde{H}_\text{I} (f) \frac{\Delta \rpf (f)}{2 \rp}, \\ \nonumber
\Delta \tilde{\mathcal{Q}}(f) & = &
		\tilde{F}_\text{Q} (f) \, \Delta \tilde{T}_\text{qp} (f)
		+ \tilde{H}_\text{Q} (f) \frac{\Delta \rpf (f)}{2 \rp}.
\end{eqnarray}
where the first term accounts for a change in the effective quasiparticle temperature and the second term for a change in the applied readout power. If the readout power is held constant, only the first term in each expression is required; if the readout power is changed, but there is no internal heating, only the second term in each expression is required; if the readout power is changed and there is internal heating, the full expressions are needed.

\section{Operating point}\label{sec:operating_temp}

To use the expressions derived in Section \ref{sec:resonator_and_readout} it is necessary to know the quiescent value of the effective quasiparticle temperature, $T_\text{qp}$, which is not the same as the bath temperature because of readout-power heating.
The analysis presented here is generic, and applies to all electrothermal models of the kind shown in Figure \ref{fig:thermal_model}.
Let $\mathbf{T}$, $\mathbf{U}$ and $\mathbf{q}$ denote state vectors of temperature, internal energy and net input power for each of the elements shown in Figure \ref{fig:thermal_model}.
For example, the first elements of the state vectors correspond, respectively, to the temperature, internal energy, and net input power $\hf{R} + \hf{S} - \hf{I}$, for the quasiparticle system.
The second elements of the state vectors correspond to the phonon system of the superconductor.
The net power flowing into each element depends on a number of external parameters -- such as the signal power, substrate heater power, readout power, etc. -- and for brevity we shall collect these variables together into a single vector denoted $\mathbf{v}$.

If the external parameters are held constant at $\mathbf{v}_0$, $\mathbf{T}$ evolves from a given starting point towards a steady-state value of $\mathbf{T}_0$ for which
\begin{equation}\label{eqn:ss_condition}
	\mathbf{q}(\mathbf{T}_0, \mathbf{v}_0) = \mathbf{0},
\end{equation}
henceforth referred to as the operating point.
In previous papers\,\cite{de2010readout, thompson2013dynamical}, we have discussed the solution of (\ref{eqn:ss_condition}) and its consequences for device behaviour, and we shall not repeat this work here.
What is found, for example, is that high readout power heating causes variation in $\mathbf{T}_0$ across a frequency sweep, accounting for the distortion of the resonance curve observed in experiments.

There can in fact be multiple values of $\mathbf{T}_0$ for which (\ref{eqn:ss_condition}) is satisfied, corresponding to different possible states of the resonator.
However, for these states to correspond to realizable operating points they must also be dynamically stable against perturbations.
The existence of two or more stable states leads to the hysteretic resonance curves seen experimentally.
In \cite{de2010readout, thompson2013dynamical} we considered the stability criterion for a system with a single heat capacity, and here we present the generalisation to systems with several heat capacities.

Consider what happens if $\mathbf{T}$ is perturbed instantaneously by $\Delta \mathbf{T}$ away from $\mathbf{T}_0$, say by noise.
Conservation of energy requires
\begin{equation}
\label{eqn:energy_conservation}
	\partial_t \Delta \mathbf{U}(t)
	=
	-\gmat \cdot \Delta \mathbf{T}(t),
\end{equation}
or purely in terms of temperature
\begin{equation}
	\partial_t \Delta \mathbf{T}(t)
	=
	-\cmat^{-1} \cdot \gmat \cdot \Delta \mathbf{T}(t).
\end{equation}
$\gmat$ is a thermal-conductance matrix with elements defined by
\begin{equation}
\label{eqn:def_g_matrix}
	\{ \gmat \}_{mn}	
	= - \left( \frac{\partial q_{m}}{\partial T_{n}}
		\right)_{\mathbf{T}_0, \mathbf{v}_0},
\end{equation}
while $\cmat$ is a heat capacity matrix with elements
\begin{equation}\label{eqn:def_c_matrix}
	\{ \cmat \}_{mn}
	= \left( \frac{\partial U_{m}}{\partial T_{n}}
		\right)_{\mathbf{T}_0, \mathbf{v}_0}.
\end{equation}
For clarity, we emphasise $q_m$, $U_m$ and $T_m$ are respectively the net input power, internal energy and temperature of the $m^\text{th}$ system, corresponding to the $m^\text{th}$ entries of $\mathbf{q}$, $\mathbf{U}$ and $\mathbf{T}$.
Both partial derivatives are evaluated at the system temperatures $\mathbf{T}_0$ at the operating point and for the assumed values, $\mathbf{v}_0$, of the external parameters.
Because the internal energy of each element depends only on its own temperature, $\cmat$ is diagonal.
The size of the perturbation of the system away from equilibrium is measured by $|\Delta \mathbf{T}(t)|^2$, which from (\ref{eqn:energy_conservation}) evolves in time according to
\begin{equation}\label{eqn:mod_temp_perturbation_de}
	\partial_t |\Delta \mathbf{T}(t)|^2
	= - 2 \Delta \mathbf{T}(t)^\dagger \cdot \smat \cdot \Delta \mathbf{T}(t),
\end{equation}
where $\smat$ is the matrix
\begin{equation}\label{eqn:def_s_matrix}
	\smat
	= \frac{1}{2} \biggl\{
		\cmat^{-1} \cdot \gmat +
		\bigl[ \cmat^{-1} \cdot \gmat \bigr]^\dagger
	\biggr\}.
\end{equation}
(\ref{eqn:mod_temp_perturbation_de}) implies that provided
\begin{equation}\label{eqn:stability_condition}
	\mathbf{x}^\dagger \cdot \smat \cdot \mathbf{x} > 0
\end{equation}
for all $\mathbf{x}$, any perturbation decreases in magnitude with time, and the state is stable.
If (\ref{eqn:stability_condition}) is not satisfied, noise processes will inevitably excite perturbations that grow in time and drive $\mathbf{T}$ into a different state.
The possible operating points of a device for given $\mathbf{v}_0$ can thus be found by checking the roots of (\ref{eqn:ss_condition}) against the stability condition of (\ref{eqn:stability_condition}).
(\ref{eqn:stability_condition}) corresponds to demanding $\smat$ is positive definite, which can be tested numerically by checking that its eigenvalues are all non-zero and positive.

It is useful to consider the physical processes included in $\gmat$, as the concepts are central to the paper. The classical definition of thermal conductance is the constant of proportionality relating differential changes in heat flow down a thermal link to changes in the temperature of its ends. These thermal conductances, derived from the various heat-flow equations listed in Section \ref{sec:thermal_model}, make up the majority of the elements in $\gmat$.
However, $\gmat$ also contains effective conductance terms not derived from thermal links.
For example, changes in the quasiparticle temperature affect the resonance curve and therefore the readout power dissipated in the device, which contributes to $\mathbf{q}$.
This process contributes an effective thermal conductance in $\gmat$, which links the quasiparticle system to a hypothetical bath.
The sign and magnitude of this effective conductance depend on the readout power and frequency, providing a mechanism for electrothermal feedback.
Both types of mechanism are intrinsic in the definition of (\ref{eqn:def_g_matrix}).

Two assumptions have crept into the analysis. In (\ref{eqn:def_c_matrix}) and (\ref{eqn:def_g_matrix}), the elements of both $\mathbf{U}$ and $\mathbf{q}$ have been assumed to be instantaneously related to temperature.
This quasistatic approach is sufficient for first-order stability analysis, but we shall reconsider the assumption regarding $\mathbf{q}$ when discussing small-signal response.

\section{Small-signal response}\label{sec:linearised_model}

Now consider how changes in external parameters, such as optical signal power, calibration heater power, readout power, etc., lead to  changes in the quasiparticle temperature. The result will be a full small-signal model of device behaviour.

\subsection{Linearising thermal response}\label{sec:linearised_thermal_response}

In the small-signal limit, the change $\Delta \mathbf{T}(t)$ in the system temperatures, in response to a perturbation $\Delta \mathbf{v} (t)$ in external parameters, can be found by linearising (\ref{eqn:energy_conservation}) around the operating point $\mathbf{T} = \mathbf{T}_0$ and $\mathbf{v} = \mathbf{v}_0$.
To first order, the internal energies of the elements vary as
\begin{equation}
\label{eqn:linearised_energy_storage}
	\mathbf{U}(t) = \mathbf{U}_0 + \int_0^\infty \Re \left[
		\cmat \cdot \Delta \tilde{\mathbf{T}}(f)
		e^{2 \pi i f t} \right] \, df,
\end{equation}
where $\mathbf{U}_0$ is the value of $\mathbf{U}$ at the operating point.
In general, some of the heat capacities may be frequency dependent, for example in the case of TLSs, but for brevity we shall not include this possibility here.
Similarly, the net energy flows into the elements vary as
\begin{equation}\label{eqn:linearised_heat_flow}
	\Delta \mathbf{q}(t) \approx \int_0^\infty
	\Re \left[ \left[ -\gmat \cdot \Delta \tilde{\mathbf{T}}(f)
		+ \rmat \cdot \Delta \tilde{\mathbf{v}}(f) \right]
		e^{2 \pi i f t} \right]
	\, df,
\end{equation}
where the thermal conductance matrix $\gmat$ is calculated using (\ref{eqn:def_g_matrix}) and the parameter responsivity matrix $\rmat$ is defined as
\begin{equation}\label{eqn:def_r_matrix}
	\{ \rmat \}_{mn} =
	\left( \frac{\partial q_m}{\partial v_n} \right)_{\mathbf{T}_0, \mathbf{v}_0},
\end{equation}
where $v_m$ is the $m^\text{th}$ external parameter, i.e. the $m^\text{th}$ entry of $\mathbf{v}$.
The elements of $\rmat$ give the rates of change of the power entering an element with respect to variations in external sources.
In (\ref{eqn:linearised_energy_storage})--(\ref{eqn:def_r_matrix}), we have again assumed the quasistatic limit where $\gmat$, $\cmat$ and $\rmat$ do not depend on $f$.
In other words, changes in $\Delta \tilde{\mathbf{T}}(f)$ and changes in $\Delta \tilde{\mathbf{v}}(f)$ occur on time scales that are long compared with the bandwidths of the processes described by $\gmat$, $\cmat$ and $\rmat$.
This assumption may not always be appropriate in the case of readout power heating, as will be discussed in the next section.

Substituting (\ref{eqn:linearised_energy_storage}) and (\ref{eqn:linearised_heat_flow}) into (\ref{eqn:energy_conservation}) and solving, we find
\begin{eqnarray}\label{eqn:thermal_response_matrix}
	\Delta \tilde{\mathbf{T}}(f)
	 & = &
	\left\{ \left[ \gmat + 2 \pi i f \cmat \right]^{-1} \! \! \cdot \rmat \right\}
	\cdot \Delta \tilde{\mathbf{v}}(f), \\ \nonumber
 	& = & \kmat(f) \cdot \Delta \tilde{\mathbf{v}}(f)
\end{eqnarray}
which describes the small-signal thermal response to variations in the independent variables.
(\ref{eqn:thermal_response_matrix}) can then be used with (\ref{eqn:total_modulation_response}) to calculate the changes in the outputs of the I/Q mixer in response to changes in external parameters.
In the case of a KID on a bulk substrate, with only quasiparticle heating, (\ref{eqn:thermal_response_matrix}) reproduces the simple model discussed previously\,\cite{thompson2013dynamical}.

\subsection{Readout power response}\label{sec:readout_power_response}

A small-signal model using the quasistatic forms for $\gmat$, $\cmat$ and $\rmat$ is sufficient when thermal processes limit the response time of the device. However in some cases it may be the electrical response time of the resonator that is the constraint.
For example, a 5\,GHz resonator with a total Q-factor in excess of $10^6$ has a response time of the order of 100\,$\mu s$, which is comparable with typical quasiparticle lifetimes\,\cite{zmuidzinas2012superconducting}.
It is then necessary to take into account the resonator dynamics and (\ref{eqn:linearised_heat_flow}) becomes
\begin{equation}\label{eqn:linearised_heat_flow_lag}
	\Delta \mathbf{q}(t) \approx \int_0^\infty
	\Re \bigl[ \bigl[ -\gmat(f) \cdot \Delta \tilde{\mathbf{T}}(f) \nonumber
		+ \rmat(f) \cdot \Delta \tilde{\mathbf{v}}(f) \bigr]
	e^{2 \pi i f t} \bigr]
	\, df,
\end{equation}
where $\gmat$ and $\rmat$ are now assumed to depend on the frequency $f$ of the perturbations.
Most of the terms in $\gmat$ and $\rmat$ can be still be calculated using the quasistatic forms of (\ref{eqn:def_g_matrix}) and (\ref{eqn:def_r_matrix}), but those associated with the readout power heating, $\hf{R}$, must be treated differently.
In the quasistatic limit these are given by
\begin{eqnarray}
\label{eqn:def_g_matrix_readout}
	G_\text{R}
	&= -\Biggl( \frac{\partial \hf{R}}{\partial T_\text{qp}}
		\Biggr)_{\mathbf{T}_0, \mathbf{v}_0} \nonumber \\
	&= -\rp \frac{\partial \left[ 1 - |S_{11} (\nu_\text{r}, T_\text{qp})|^2
	- |S_{21}  (\nu_\text{r}, T_\text{qp})|^2 \right]}{\partial T_\text{qp}},
\end{eqnarray}
and
\begin{eqnarray}
\label{eqn:def_r_matrix_readout}
	R_\text{R}	
	&= \Biggl( \frac{\partial \hf{R}}{\partial \rp}
		\Biggr)_{\mathbf{T}_0, \mathbf{v}_0} \nonumber \\
	&=  1 - |S_{11} (\nu_\text{r}, T_\text{qp})|^2
	- |S_{21} (\nu_\text{r}, T_\text{qp})|^2.
\end{eqnarray}
In this section we will develop replacement expressions that take full account of the finite response time of the resonator.
In what follows, we will suppress the temperature dependence of the $S$-parameters for notational convenience and the operating-point values should be assumed.

First consider $G_\text{R}$. The instantaneous power dissipated in the resonator by the readout signal is given by $\hf{R}(t) = R(t) I^2(t)$.
When the resistance of the equivalent circuit is perturbed, the current also changes and $q_\text{R} (t) = \left[ R_0 + \Delta R(t) \right] \left[ I(t) + \Delta I(t) \right]^{2}$, and so to first-order
\begin{equation}\label{eqn:inst_power}
	\Delta \hf{R}(t) \approx  \Delta R(t) I^2(t) +  2 R_0  I(t) \Delta I(t).
\end{equation}
Substituting the time-dependent quantities in terms of their spectra, and ignoring fluctuations in power that occur at frequencies greater than the readout frequency gives
\begin{eqnarray}\label{eqn:inst_power_spec}
	\Delta q_{R}(t) & \approx & \frac{|I_0|^2}{2} \int_{0}^{\nu_\text{r}} \Re \left[ \Delta \tilde{R} (f) e^{2 \pi i f t} \right] \, df \\ \nonumber
 & + & R_0 \int_{0}^{\nu_\text{r}} \Re \left[ I_0 \Delta \tilde{I}^{\ast} ( \nu_\text{r} - f ) e^{2 \pi i f t} \right] \, df \\ \nonumber
 & + & R_0 \int_{0}^{\nu_\text{r}} \Re \left[ I_0^{\ast} \Delta \tilde{I} ( \nu_\text{r} + f ) e^{2 \pi i f t} \right] \, df.
\end{eqnarray}
The first term arises from the time-varying resistance modulating the average value of the readout current, and it is not limited by the bandwidth of the resonator.
The second and third terms correspond, respectively, to upper and lower sidebands resulting from intermodulation between the steady state and perturbation currents.

(\ref{eqn:ss_resonator_current}), (\ref{eqn:u_pertubration_current}) and (\ref{eqn:resonator_s11}), together with the result
\begin{equation}\label{eqn:resistance}
	Z_\text{R} (\nu) = \frac{S_{21} (\nu)}{2 \left[ 1 - S_{21} (\nu) \right]} Z_0,
\end{equation}
can be used to rewrite (\ref{eqn:inst_power_spec}) in terms of $S$-parameters and temperature perturbations.
Casting the result into the form of (\ref{eqn:linearised_heat_flow_lag}), it can be shown, after some tedious algebra, that
\begin{eqnarray}\label{eqn:GR}
	&G_\text{R} (f) = \nonumber \\
	&  -\rp \left\{
		\frac{ |S_{11}(\nu_\text{r})|^2
		+ \alpha
		S_{11}(\nu_\text{r} + f)}{S_{11}(\nu_\text{r})^2} \,
		\frac{\partial S_{21}(\nu_\text{r})}{\partial T_{qp}} \right. \nonumber \\ 
 	& + \left.
 		\frac{ |S_{11}(\nu_\text{r})|^2
 		+ \alpha
 		S_{11}^{\ast}(\nu_\text{r} - f)}{\left[ S_{11}^\ast(\nu_\text{r}) \right]^2} \,
 		\frac{\partial S_{21}^{\ast}(\nu_\text{r})}{\partial T_{qp}} \right\},
\end{eqnarray}
where $\alpha =  1 - |S_{11}(\nu_\text{r})|^2 - |S_{21}(\nu_\text{r})|^2$.
Although (\ref{eqn:GR}) is complicated, it is easy to evaluate numerically, and it accounts for the frequency dependence of the absorbed readout power when the quasiparticle temperature is modulated.
Notice that the terms  $S_{11}(\nu_\text{r} + f)$ and $S_{11}^{\ast}(\nu_\text{r} - f)$, which involve the readout frequency, may be different when the modulation frequency is comparable with the bandwidth of the resonator.
In particular, $G_\text{R}$ can have an imaginary component, leading to the possibility of an oscillatory feature in the thermal response as described by (\ref{eqn:thermal_response_matrix}) and indeed the outputs of the I/Q mixer given by (\ref{eqn:temperature_responsivity}).
This feature corresponds to the exchange of energy stored in the electromagnetic fields of the resonator and energy stored in the heat capacity of the thermal elements.

If $f$ is sufficiently small such that $S_{21} ( \nu_{r} \pm f) \approx S_{21} ( \nu_{r})$,
(\ref{eqn:GR}) gives
\begin{equation}\label{eqn:GR_limit1}
	G_\text{R} (f \rightarrow 0) \approx
	-2 \rp \Re \left\{ \left[ 1 - 2 S_{21}^{\ast}(\nu_\text{r}) \right] \frac{\partial S_{21}(\nu_\text{r})}{\partial T_{qp}} \right\},
\end{equation}
which is identical to the quasistatic limit found by evaluating the partial derivative in (\ref{eqn:def_g_matrix_readout}).
In this case the response function is real, and oscillatory interactions between the electrical and thermal systems do not occur.
In the case where $f$ exceeds the bandwidth of the resonator  $S_{11}(\nu_\text{r} + f) \rightarrow 0$ and $S_{11}^{\ast}(\nu_\text{r} - f) \rightarrow 0$, because the resonator is an open circuit out of band.
(\ref{eqn:GR}) then becomes
\begin{equation}\label{eqn:GR_limit2}
	G_\text{R} (f \rightarrow \infty)
	\approx -2 \rp \Re \left\{ \left[ \frac{S_{21}^{\ast}(\nu_\text{r})-1}{S_{21}(\nu_\text{r})-1} \right]
	\frac{\partial S_{21}(\nu_\text{r})}{\partial T_{qp}} \right\}.
\end{equation}
This term comes from the first term in (\ref{eqn:inst_power_spec}) and it remains even for very high modulation frequencies, showing that feedback can occur for modulation frequencies much greater than the line width of the resonator.
Physically, the effect is due to the resistance, and therefore the power, being modulated by the quasiparticle temperature, even though the currents cannot change sufficiently quickly.
In both (\ref{eqn:GR_limit1}) and (\ref{eqn:GR_limit2}), feedback only occurs when the readout frequency lies within the width of the resonance, as expected.

Now consider $R_\text{R}(f)$, which describes how the dissipated power responds to changes in the applied readout power.
In this case, the equivalent resistance of the resonator is constant at $R_0$, and (\ref{eqn:inst_power_spec}) immediately gives the dissipated power as
\begin{eqnarray}\label{eqn:inst_power_pert}
	\Delta \hf{R}(t) & \approx &  R_0 \int_{0}^{\nu_\text{r}} \Re \left[ I_0 \Delta \tilde{I}^{\ast} ( \nu_\text{r} - f ) e^{2 \pi i f t} \right] \, df \\ \nonumber
 & + & R_0 \int_{0}^{\nu_\text{r}} \Re \left[ I_0^{\ast} \Delta \tilde{I} ( \nu_\text{r} + f ) e^{2 \pi i f t} \right] \, df.
\end{eqnarray}
Using  (\ref{eqn:ss_resonator_current}), (\ref{eqn:modulation_term}) and (\ref{eqn:resistance}) one finds that
\begin{eqnarray} \label{eqn:resonator_qr_m}
	& \Delta \hf{R}(t) =   4 P_r \Re \left[ S_{21}(\nu_{r}) - | S_{21}(\nu_{r}) |^{2} \right] \\ 	\nonumber
 	& \times \int_{0}^{\nu_\text{r}} \Re \left\{ \left[ \frac{S_{11}( \nu_{r}+f )}{S_{11} (\nu_{r})}
 	+ \frac{S_{11}^{\ast}( \nu_{r}-f )}{S_{11}^{\ast} (\nu_{r})} \right] \tilde{m}(f) e^{2 \pi i f t} \right\} \, df,
\end{eqnarray}
and using (\ref{eqn:modulation}) gives
\begin{eqnarray} \label{eqn:resonator_qr_power}
	&\Delta \hf{R}(t) = \frac{1}{2} \left[ 1 - |S_{11}(\nu_\text{r})|^{2} - |S_{21}(\nu_\text{r})|^{2} \right] \nonumber \\
	& \times \int_{0}^{\nu_\text{r}} \Re \left\{ \left[ \frac{S_{11}( \nu_{r}+f )}{S_{11} (\nu_{r})} + \frac{S_{11}^{\ast}( \nu_{r}-f )}{S_{11}^{\ast} (\nu_{r})} \right] \Delta P_r(f) e^{2 \pi i f t} \right\} \, df,
\end{eqnarray}
and then according to (\ref{eqn:linearised_heat_flow_lag})
\begin{eqnarray}
	R_\text{R}(f) = & \frac{1}{2}
		\left[ 1 - |S_{21}(\nu_\text{r})|^{2} - |S_{11}(\nu_\text{r})|^{2} \right] \nonumber \\
	& \times
		\left[ \frac{S_{11}( \nu_{r}+f )}{S_{11} (\nu_{r})}
		+ \frac{S_{11}^{\ast}( \nu_{r}-f )}{S_{11}^{\ast} (\nu_{r})} \right].
	\label{eqn:resonator_r}
\end{eqnarray}
In the quasistatic limit, $f \rightarrow 0$, (\ref{eqn:def_r_matrix_readout}) is recovered.
In the case where $\nu_\text{r} \pm f$ moves outside of the resonance, $R_\text{R} (f) \rightarrow 0$.

\section{Overall small-signal response}\label{sec:overall_small_signal_responsivity}

(\ref{eqn:thermal_response_matrix}) can be used to calculate all of the temperature variations in the system when any of the external parameters is varied. According to (\ref{eqn:temperature_responsivity}), it is only necessary to know the change in quasiparticle temperature to calculate the outputs of the I/Q mixer.
Thus, we take the appropriate row $\mathbf{k}_0(f)$ of $\kmat(f)$ corresponding to the quasiparticle temperature, and write
\begin{eqnarray}\label{eqn:full_responsivity}
	\Delta \tilde{\mathcal{I}}(f) & = &
		\tilde{F}_\text{I} (f) \, \mathbf{k}_0 (f) \cdot \Delta \tilde{\mathbf{v}}(f)
		+ \tilde{H}_\text{I} (f) \frac{\Delta \rpf (f)}{2 \rp}, \\ \nonumber
	\Delta \tilde{\mathcal{Q}}(f) & = &
		\tilde{F}_\text{Q} (f) \, \mathbf{k}_0 (f) \cdot \Delta \tilde{\mathbf{v}}(f)
		+  \tilde{H}_\text{Q} (f) \frac{\Delta \rpf (f)}{2 \rp} .
\end{eqnarray}
(\ref{eqn:full_responsivity}) is the primary result of this paper, because it provides a full description of the small-signal response measured at the mixer outputs.
It includes the electrothermal feedback effects associated with the readout power, and also effects due the potentially narrow filtering of high-Q resonators when the readout power is varied.

\section{Noise}\label{sec:noise}

Fast temporal fluctuations appear in the outputs of the I/Q mixer as a consequence of noise internal to the KID, and noise associated with external sources such as the readout electronics\,\cite{zmuidzinas2012superconducting}.
When modelling low-noise detectors, such as bolometers, internal noise is usually treated by considering the physical origins of each of the microscopic processes, and then incoherently adding the contributions.
The overall result usually turns out to be an instance of the more general fluctuation-dissipation theorem, where the noise is determined by the mean relaxation behaviour of the complete device rather than the exact statistics of the individual processes.
For complex electrothermal systems, where thermal fluctuation noise may occur as a consequence of energy being exchanged randomly between elements, for example through the Kapitza conductance and weak quasiparticle-phonon coupling, it is beneficial to adopt a general approach based on Langevin analysis\,\cite{montroll2012fluctuation}.

Our analysis is carried out in two steps: (i) Introduce a set of equivalent noise-power sources that drive energy into each of the elements of the electrothermal model, corresponding to fluctuations in the extensive variables of the system, most notably $\mathbf{U}$.
(ii) Use a thermodynamic argument to determine the magnitude of the thermal fluctuations in $\mathbf{U}$, and thereby determine the corresponding magnitudes of the noise-power inputs.
The advantage of this approach is that system noise can be calculated for any arbitrary arrangement of elements.
Also, our scheme opens the door to more general non-equilibrium calculations\,\cite{irwin2006thermodynamics}.
By way of verification, it will be shown in Section \ref{sec:gr_noise} that when quasiparticle-phonon coupling is the dominant weak thermal link, the scheme reproduces generation-recombination noise without any explicit reference to the physics of this process.

Connect a set of randomly varying power sources to the thermal elements of Figure  \ref{fig:thermal_model}.
The sources are included mathematically by expressing (\ref{eqn:energy_conservation}) in terms of the variations in internal energy around the quiescent value, and adding to it the noise-power vector $\Delta \mathbf{P}_\text{N}(t)$:
\begin{equation}\label{eqn:u_fluctuation_de}
	\partial_t \Delta \mathbf{U}(t) =
	-\gmat \cdot \cmat^{-1} \cdot \Delta \mathbf{U}(t)
	+ \Delta \mathbf{P}_\text{N}(t).
\end{equation}
Following the usual procedure for Langevin analysis, assume the following: (i) The added noise has zero mean,
\begin{equation}\label{eqn:noise_power_zero_mean}
	\langle \mathbf{P}_\text{N}(t) \rangle = \mathbf{0},
\end{equation}
which occurs because if this were not the case the operating point would simply shift until it is satisfied.
(ii) The second order correlations satisfy
\begin{equation}\label{eqn:noise_power}
	\langle \Delta \mathbf{P}_\text{N}^{} (t_1)
	\Delta \mathbf{P}_\text{N}^{\dagger} (t_2)  \rangle
	= \mathsf{N} \delta (t_1 - t_2),
\end{equation}
where $\mathsf{N}$ is a real-valued symmetric matrix.
The physical motivation for this assumption is that the microscopic processes driving the noise are only correlated only on very short time scales.
(iii) All higher order correlation functions are zero.
More generally, these assumptions can be argued on the basis that thermodynamics predicts that the fluctuations in $\mathbf{U}$ form a Gaussian process, and (i)--(iii) correspond to the noise process with the fewest free-parameters that can reproduce this effect.
If $\Delta \mathbf{U}$ has the value $\Delta \mathbf{U}_0$ at time $t_0$, the general solution of (\ref{eqn:u_fluctuation_de}) for $t > t_0$ is
\begin{eqnarray}\label{eqn:u_general_solution}
	\Delta \mathbf{U}(t) &=& e^{-(t - t_0) \gmat^{} \cdot \cmat^{-1}}
	\! \! \! \cdot \Delta \mathbf{U}_0 \nonumber \\
	&& + \int_{t_0}^{t} e^{-(t' - t_0) \gmat^{} \cdot \cmat^{-1}}
	\! \! \! \cdot \Delta \mathbf{P}_\text{N} (t') \, dt',
\end{eqnarray}
where use has been made of the exponential of a matrix\,\cite{riley2006mathematical}.
Using (\ref{eqn:noise_power}), the equal-time correlation matrix of the fluctuations is
\begin{eqnarray}\label{eqn:u_variance}
	&\langle \Delta \mathbf{U}(t)^{} \Delta \mathbf{U}(t)^\dagger \rangle
	\nonumber \\
	&= e^{-(t - t_0) \gmat^{} \cdot \cmat^{-1}}
	\! \! \! \cdot \Delta \mathbf{U}_0^{} \Delta \mathbf{U}_0^{\dagger}
	\cdot e^{-(t - t_0) [ \gmat^{} \cdot \cmat^{-1} ]^\dagger}
	\nonumber \\
	& + \int_{t_0}^{t} e^{-(t' - t_0) \gmat^{} \cdot \cmat^{-1}}
	\cdot \mathsf{N} \cdot
	e^{-(t' - t_0) [ \gmat^{} \cdot \cmat^{-1} ]^\dagger}
	\, dt'.
\end{eqnarray}
The first term describes the relaxation of the system back to equilibrium from the starting condition, and the second term describes the steady-state fluctuations.
In the limit $t >> t_0$, the operator $\exp[-(t - t_0) \gmat^{} \cdot \cmat^{-1}]$ must tend to zero if $\mathbf{T}_0$ is an equilibrium point, as discussed in Section~\ref{sec:operating_temp}.
Evaluating the integral in (\ref{eqn:u_variance}) then gives
\begin{eqnarray}\label{eqn:u_variance_long_times}
	\langle \Delta \mathbf{U}(t)^{} \Delta \mathbf{U}(t)^\dagger \rangle
	= \mathsf{M},
\end{eqnarray}
where
\begin{equation}\label{eqn:def_n_matrix}
	\mathsf{N} = \gmat^{} \cdot \cmat^{-1} \cdot \mathsf{M}
	+ \mathsf{M} \cdot \left[ \gmat^{} \cdot \cmat^{-1} \right]^\dagger.
\end{equation}
If the two-time correlation function is evaluated, the noise in $\Delta \mathbf{U}$ is found to be non-white, and the noise is correlated on timescales determined by the heat capacities and weak thermal links.
(\ref{eqn:def_n_matrix}) and (\ref{eqn:u_variance_long_times}) provide a way of calculating $\mathsf{N}$ from the observed fluctuations in $\mathbf{U}$, but we
have said nothing about their magnitudes.

A standard result in thermodynamics\,\cite{adkins1983equilibrium} is that the internal energy of a heat capacity $C$ in equilibrium with a bath at temperature $T$ exhibits thermal fluctuations of magnitude
\begin{equation}\label{eqn:u_fluctuations}
	\langle \Delta U^2 \rangle = k_\text{b} T^2 C.
\end{equation}
Strictly, the assumption of thermal equilibrium is not satisfied for an operating point above the bath temperature, but this is a standard assumption in bolometer theory.
The question of noise in detectors under non-equilibrium conditions is an open one, and a number of different approaches have been proposed\,\cite{mather1982bolometer,irwin2006thermodynamics}.
The scheme presented here, would lend itself to non-equilibrium calculations.

In the spirit of equilibrium thermodynamics, assume that the fluctuations in the energy of each element are given by (\ref{eqn:u_fluctuations}), but evaluated at the operating temperature of that element, so
\begin{equation}\label{eqn:non_equlibrium_u_fluctuations}
	\langle \Delta \mathbf{U}(t)^{} \Delta \mathbf{U}(t)^\dagger \rangle
	= k_\text{b} \cmat \cdot \mathsf{T}_0 \cdot \mathsf{T}_0,
\end{equation}
where $\{\mathsf{T}_0\}_{mn} = \{ \mathbf{T}_0 \}_m \delta_{mn}$.
Inserting (\ref{eqn:non_equlibrium_u_fluctuations}) into (\ref{eqn:def_n_matrix}), we find
\begin{equation}\label{eqn:n_matrix}
	\mathsf{N} = k_\text{b} \left[
		\gmat^{} \cdot \mathsf{T}_0 \cdot \mathsf{T}_0
		+ \mathsf{T}_0 \cdot \mathsf{T}_0 \cdot \gmat^\dagger
	\right].
\end{equation}
We can calculate the, spatial, correlations between the noise sources, from the symmetric part of the thermal-conductance matrix $\gmat$, which is a fluctuation-dissipation relation for KIDs.
(\ref{eqn:n_matrix}) is closely related to the usual practice of placing noise sources across thermal conductances to represent noise.

In order to determine the recorded noise it is necessary to propagate the effects of the equivalent noise sources to the outputs of the I/Q-mixer.
Comparing (\ref{eqn:linearised_heat_flow}) and (\ref{eqn:u_fluctuation_de}), we see that the noise power term must enter (\ref{eqn:linearised_heat_flow}) in the same way as variations in the external parameters term.
However, we need the frequency domain representation of $\Delta \mathbf{P}_\text{N}(t)$.
Taking the Fourier transforms of (\ref{eqn:noise_power_zero_mean}) and (\ref{eqn:noise_power}), we obtain
\begin{equation}\label{eqn:freq_domain_zero_mean}
	\langle \Delta \tilde{\mathbf{P}}_N(f) \rangle = \mathbf{0},
\end{equation}
and
\begin{equation}\label{eqn:freq_domain_correlation}
	\langle \Delta \tilde{\mathbf{P}}_N^{} (f_1)
	\Delta \tilde{\mathbf{P}}_N^\dagger (f_2) \rangle
	= 2 \mathsf{N} \delta (f_1 - f_2),
\end{equation}
for single-sided spectra.
Replacing $\rmat \cdot \Delta \tilde{v}(f)$ with $\Delta \tilde{\mathbf{P}}_\text{N}(f)$ in the subsequent analysis in Section \ref{sec:linearised_model}, we find the corresponding noise signal has the properties
\begin{equation}\label{eqn:iq_mean_noise}
	\langle \tilde{A}(f) \rangle = 0
\end{equation}
and
\begin{eqnarray}\label{eqn:iq_noise_correlations}
	\langle \tilde{A}(f_1) \tilde{B}(f_2)  \rangle =&
		2 \tilde{F}_A (f_1) \, \hat{\mathbf{x}} \cdot
		\left[ \gmat^{} + 2 \pi i f_1 \cmat \right]^{-1} \nonumber \\
	&\cdot \mathsf{N} \cdot
		\left[ \gmat^\dagger - 2 \pi i f_2 \cmat \right]^{-1} \! \! \cdot
		\hat{\mathbf{x}}^\dagger \, \tilde{F}_B (f_2).
\end{eqnarray}
where $A$ and $B$ can be $I$ or $Q$ and $\hat{\mathbf{x}}$ is a unit vector with a non-zero entry at the element of $\Delta \tilde{\mathbf{T}} (f)$ corresponding to the temperature of the quasiparticle system.

Of particular interest is the Noise Equivalent Power (NEP), which requires explicit knowledge of the how the KID is read out, i.e. whether the transmission phase-shift, amplitude-shift or some weighted linear combination of the two is used.
The small-signal transformations discussed in (\ref{sec:overall_small_signal_responsivity}) and (\ref{eqn:iq_noise_correlations}) can be used, respectively, to calculate the appropriate responsivity and noise, which then allows the calculation of NEP.

\section{Comparison with other models}\label{sec:model_equivalence}

The usual approach to modelling KIDs is to use a simplified quasiparticle-number model to understand small-signal behaviour, and non-linear resonator theory to explain hysteresis\,\cite{zmuidzinas2012superconducting,swenson2013operation}.
In this section we show how our more general electrothermal model reproduces the behaviour of these other approaches, under a single framework, in the appropriate limits.

\subsection{Quasiparticle-number model}\label{sec:responsivity_equivalence}

The quasiparticle-number model assumes that all of the microscopic systems are at the bath temperature, and there exists an excess of quasiparticles over the expected thermal population\,\cite{zmuidzinas2012superconducting}.
The total number of quasiparticles, $N_\text{qp}$, then parameterizes the excited state.
The corresponding reduced form of our model assumes that only the quasiparticle temperature $T_\text{qp}$ can vary, and that
\begin{equation}\label{eqn:qp_bolometer}
	\left[ C_0 \, \partial_t + G_0 \right] \Delta T_\text{qp}(t)
	= \Delta P_\text{qp}(t),
\end{equation}
where $G_0$ is the effective thermal conductance between the quasiparticles and a fictitious bath at the operating temperature, $C_0$ is the heat capacity of the quasiparticle system and $\Delta P_\text{qp}$ is the change in the net power flow into the quasiparticle system from external sources.
(\ref{eqn:qp_bolometer}) is effectively a treatment of a simple KID as a classical bolometer, where the quasiparticle system functions as the isolated heat capacity whose temperature is monitored.

At low reduced temperatures, $N_\text{qp}$ and $U_\text{qp}$, the internal energy of the quasiparticle system, are approximately
\begin{eqnarray}
\label{eqn:qp_number}
	N_\text{qp} = 2 n_0 V \sqrt{2 \pi k_\text{b} T \scgap} e^{-\scgap / k_\text{b} T}  \\
\label{eqn:qp_energy}
	U_\text{qp} = \scgap N_\text{qp},
\end{eqnarray}
as is shown in \ref{sec:approx_qp_values}.
The temperature dependence of $\scgap$ is sufficiently weak that it can be ignored to first-order.
Taking the temperature derivatives of (\ref{eqn:qp_number}) and (\ref{eqn:qp_energy}) then gives
\begin{eqnarray}
\label{eqn:qp_fluctuation_relation}
	\Delta N_\text{qp} =
		\left( \frac{N_\text{qp} \scgap}{k_\text{n} T_\text{qp}^2} \right)
		\Delta T_\text{qp} \\
\label{eqn:qp_heat_capacity}
	C_0 = \frac{N_\text{qp} \scgap^2}{k_\text{n} T_\text{qp}^2},
\end{eqnarray}
which relate changes in quasiparticle number to changes in temperature.
Substituting (\ref{eqn:qp_fluctuation_relation}) and (\ref{eqn:qp_heat_capacity}) into (\ref{eqn:qp_bolometer}), we recover the usual rate equation for the excess quasiparticle number in the quasi-particle number model,
\begin{equation}\label{eqn:qp_model_governing_equation}
	\left[ \partial_t + \tau^{-1} \right] \Delta N_\text{qp}(t) = \Gamma(t),
\end{equation}
where the relaxation time $\tau$ and generation rate $\Gamma$ are related to the thermal parameters by
\begin{eqnarray}
	\label{eqn:qp_lifetime}
	\tau = C_0/ G_0 \\
	\label{eqn:qp_generation_rate}
	\Gamma = \Delta P_\text{qp}(t) / \scgap.
\end{eqnarray}

\subsection{Generation-recombination noise}\label{sec:gr_noise}

$N_\text{qp}$ is maintained by a balance of generation processes, where a Cooper pair is broken by a phonon or photon, and loss processes, where quasiparticles recombine into Cooper pairs or are lost from the superconductor.
These processes are intrinsically random and so $N_\text{qp}$ fluctuates\,\cite{wilson2004quasiparticle}, leading to so-called Generation-Recombination (GR) noise\,\cite{zmuidzinas2012superconducting}.

GR noise arises naturally in our model as the Langevin noise source associated with the quasiparticle-phonon coupling conductance.
Following the methods of Section \ref{sec:noise}, it can be accounted for by considering a noise power input $\Delta P_\text{N}$ in (\ref{eqn:qp_bolometer}), i.e.
\begin{equation}\label{eqn:noisy_qp_governing_equation}
	\Delta P_\text{qp} (t) = \eta \Delta P_\text{S} (t) + \Delta P_\text{N} (t),
\end{equation}
where $\eta \Delta P_\text{S}(t)$ is the optical signal and the noise term satisfies
\begin{eqnarray}
	\langle \Delta P_\text{N} (t) \rangle = 0 \nonumber \label{eqn:qp_noise_mean}\\
	\langle \Delta P_\text{N} (t_1) \Delta P_\text{N} (t_2) \rangle
	= 2 k_\text{b}^{} T_\text{qp}^2 G_0^{} \delta (t_1 - t_2) \label{eqn:qp_noise_ac}.
\end{eqnarray}
The NEP is the square root of the single-sided spectral density of an equivalent white-noise input signal $\Delta P_\text{S}(t)$ that
represents the noise generated in the device. It follows from (\ref{eqn:noisy_qp_governing_equation}) and (\ref{eqn:qp_noise_ac}) that
\begin{equation}\label{eqn:gr_nep_thermal}
	\text{NEP} = \frac{1}{\eta} \sqrt{4 k_\text{b}^{} T_\text{qp}^2 G_0^{}}.
\end{equation}
(\ref{eqn:qp_heat_capacity}) can then be used with (\ref{eqn:qp_lifetime}) to express (\ref{eqn:gr_nep_thermal}) in the form
\begin{equation}\label{eqn:gr_nep_microscopic}
	\text{NEP} = \frac{2 \scgap}{\eta} \sqrt{\frac{N_\text{qp}}{\tau}}.
\end{equation}
which is the standard result for the GR-noise limited NEP of a KID \,\cite{de2012generation, zmuidzinas2012superconducting}.

An alternative line of reasoning is to use (\ref{eqn:qp_energy}) to relate changes in $U_\text{qp}$ to changes in $N_\text{qp}$.
In conjunction with (\ref{eqn:qp_heat_capacity}), this allows the conversion of the thermal fluctuations in the energy in the quasiparticle system, as described by (\ref{eqn:u_fluctuations}), into equivalent number fluctuations.
The result is
\begin{equation}\label{eqn:qp_number_fluctuations}
	\langle \Delta N_\text{qp}^{2} \rangle = N_\text{qp},
\end{equation}
which is the standard expression for the mean-square number fluctuations associated with GR noise\,\cite{de2012generation}.

\subsection{Hysteretic switching}\label{sec:hysteresis_equivalence}

The only area where the relationship with other models is not entirely clear relates to the hysteretic switching seen in resonance curves at high readout powers.
In our model, hysteretic switching arises naturally as a consequence of the quasiparticle temperature having two stable states, and the system being driven between these states by a dynamical feedback mechanism associated with the absorption of readout power.
A different, but related, approach is to consider the magnetic-field dependence of the kinetic inductance\,\cite{pippard1953experimental}.
In the context of KIDs, this effect has been treated by Swenson\,\cite{swenson2013operation} by assuming that the lumped inductance depends on the square of the current:
\begin{equation}\label{eqn:non_linear_L}
	L(I) = L \bigl[ 1 + (I / I_*)^2 + \cdots \bigr].
\end{equation}
The non-linear terms in (\ref{eqn:resonator_de}) result in Duffing-like resonator dynamics, with the possibility of multiple amplitude states for a given drive signal.
Hysteresis results when the resonator switches between these states.

Our scheme also predicts that $L$ will depend on $I^2$, but because of readout power heating.
Each of the lumped resonator parameters $R$, $L$ and $C$ depends on $T_\text{qp}$, which in turn depends on the power $P$ dissipated in the quasiparticle system.
For a given parameter $X$, we may therefore approximate
\begin{equation}
	X \approx X(T_B) + \frac{\partial X}{\partial T_\text{qp}} \, \frac{\partial T_\text{qp}}{\partial P} \, P,
\end{equation}
where the derivatives are evaluated at the bath temperature $T_\text{b}$.
$P$ can be expressed as $Z_R I^2 / 2$, where $Z_R$ is the real part of the resonator's impedance at the readout frequency.
With this substitution we can obtain an expression for each $X$ of the form of (\ref{eqn:non_linear_L}), giving an effective $I_*$, which characterizes the current at which nonlinear effects become important for each of the lumped parameters.
Consequently, we expect our model to produce similar results to that of Swenson when the non-linear terms in the reactive response ($L$) dominate, but to see differences if the non-linearity in the dissipative response ($R$) is also significant.

Ultimately, it comes down to whether one thinks that magnetic field or heating gives rise to the non-linearity that drives resonance curve distortion and bifurcation for a particular device.
The physical process must be found through experiment, and a comparison of measured and calculated values of $I_*$ may provide a way of distinguishing mechanisms.

\section{Conclusions}

We have presented an electro-thermal model of KIDs. Our approach has several advantages: (i) It takes into account readout-power heating, and properly accounts for the fact that the quasiparticle temperature is elevated above that of the bath.
(ii) It calculates the small-signal and noise behaviour as perturbations about the actual quasiparticle temperature, rather than assuming that the quasiparticle system is at the bath temperature.
(iii) It includes electrothermal feedback caused by readout power modulation, taking into account the ring-down time of the resonator.
(iv) It can be applied to complicated thermal arrangements, such as resonators on membranes, taking into account thermal fluctuation noise between the various elements.
(v) Generation-recombination noise appears naturally as a thermodynamic fluctuation of the energy stored in the quasiparticle system.
(v) It can account for the nonlinearities that give rise to resonance curve distortion and hysteretic switching, and leads to different responsivities and NEPs on the different hysteretic branches.
The scheme reduces to the commonly used quasiparticle-number model in the appropriate limits.

\section*{References}
\bibliographystyle{iopart-num}
\bibliography{thomas2014electrothermal_references}

\appendix

\section{Approximate expressions for the quasiparticle density and internal energy of the quasiparticle system}\label{sec:approx_qp_values}

(\ref{eqn:qp_number}) is frequently used to approximate the quasiparticle number density in a superconductor at low reduced temperature ($T / T_\text{C} < 0.1$), with Eisenmenger\,\cite{eisenmenger1981nonequilibrium} usually cited as the source.
Here we outline a derivation of the result, while also extending the analysis to consider the total internal energy of the quasiparticle system.
In BCS theory\,\cite{bardeen1957theory}, the total number $N_\text{qp}$ of quasiparticles and the total energy $U_\text{qp}$ of the quasiparticle system are given by
\begin{eqnarray}
	N_\text{qp} = 4 n_0 V \int_{\scgap}^\infty \frac{E}{\sqrt{E^2 - \Delta^2}}
	f(E, T) \, dE \label{eqn:total_qp_density} \\
	U_\text{qp} = 4 n_0  V\int_{\scgap}^\infty \frac{E^2}{\sqrt{E^2 - \Delta^2}}
	f(E, T) \, dE . \label{eqn:total_energy_density}
\end{eqnarray}
where $n_0$ is the number density of states at the Fermi surface of the normal state, $V$ is the volume of the superconductor, $\scgap$ is the superconducting gap energy and $f(E, T)$ is the Fermi-Dirac distribution.
At low reduced temperatures, $\scgap \gg k_\text{b} T$ and $f(E, T) \approx \exp(-E / k_\text{b} T)$.
The substitution $E = \Delta \cosh u$ can then be used to bring integrals (\ref{eqn:total_qp_density}) and (\ref{eqn:total_energy_density}) into the forms
\begin{eqnarray}
	N_\text{qp} = 4 n_0 V K_1(\scgap / k_\text{b} T) \, \scgap
	\label{eqn:approx_nqp_bessel} 	\\
	U_\text{qp} = 2 n_0 V
	\bigl[ K_0(\scgap/ k_\text{b} T) + K_2(\scgap / k_\text{b} T) \bigr] \, \scgap,
	\label{eqn:approx_uqp_bessel}	
\end{eqnarray}
where $K_n(x)$ is the $n^{th}$ modified Bessel function of the second kind.
The $K_n(x)$ have the same asymptotic form for large $x$ for all $n$ \cite{abramowitz1970handbook}, as given by $K_n(x) \approx \sqrt{\pi / 2 x} \exp(-x)$.
This approximation can be made in (\ref{eqn:approx_nqp_bessel}) and (\ref{eqn:approx_uqp_bessel}), in which case (\ref{eqn:qp_number}) and (\ref{eqn:qp_heat_capacity}) are obtained.

\section{Power flow between the quasiparticle and phonon systems}\label{sec:qp_ph_heat_flow}

In Section \ref{sec:model_equivalence}, we demonstrated a link between the relaxation time of the excess quasiparticle population and the thermal conductance.
Consider the case where the dominant relaxation mechanism is the heat flow $\hf{I}$ from the quasiparticles to the phonon system of the superconductor.
The relaxation time will then be $\tau = \tau_\text{r} / 2$, where $\tau_\text{r}$ is the recombination lifetime of a single quasiparticle.
The factor of a half arises because two quasiparticles are lost in each recombination event, so the quasiparticle number decays at twice the rate of a single quasiparticle considered in isolation. 
Kaplan\,\cite{kaplan1976quasiparticle} has shown $\tau_\text{r}$ can be approximated by
\begin{equation}\label{eqn:kaplan_time}
	\tau_\text{r} =
		\frac{\tau_0}{\sqrt{\pi}}
		\left( \frac{k_\text{b} T_\text{C}}{2 \zscgap} \right)^{\frac{5}{2}}
		\left( \frac{T_\text{C}}{T} \right)^{\frac{1}{2}} e^{\zscgap / k_\text{b} T_\text{qp}},
\end{equation}
where $T_\text{C}$ is the critical temperature, $\zscgap$ is the value of $\scgap$ at absolute zero and $\tau_0$ is a material-dependent lifetime.
Values of $\tau_0$ are given in Table I of \cite{kaplan1976quasiparticle}.
Using (\ref{eqn:qp_lifetime}) and (\ref{eqn:qp_heat_capacity}), we then have that the effective thermal conductance is
\begin{equation}\label{eqn:qp_ph_thermal_conductance}
	G_0=
	\frac{343 \pi k_\text{b} n_0 V \Delta_0}{4 \tau_0}
	\frac{\partial}{\partial T} \left( T e^{-2 \zscgap / k_\text{b} T} \right)_{T_\text{qp}},
\end{equation}
where we have used the BCS result $4 \Delta_0  \approx 7 k_\text{b} T_C$.
The total heat flow out of each system is found by integrating (\ref{eqn:qp_ph_thermal_conductance}) with respect to temperature, and the net heat flow follows from the difference of the two terms.
The integration can be performed trivially and we obtain a result of the form of (\ref{eqn:qp_ph_heat_flow}), as $\scgap \approx \zscgap$ at low reduced temperatures.
Further, as we are assuming $\hf{I}$ is dominated by recombination processes, $\eta_{2 \scgap (P_\text{abs})} \approx 1$ and we can make the identification $\Sigma_\text{s} \approx 343 \pi k_\text{b} n_0 \Delta_0 / 4 \tau_0 $.
Calculation of $\eta_{2 \scgap (P_\text{abs})}$ generally requires detailed simulations of both recombination and elastic scattering processes\,\cite{goldie2013non}.

\end{document}